\documentclass[preprint,aps]{revtex4}

\usepackage{graphicx}
\usepackage{dcolumn}
\usepackage{bm}
\usepackage{float,epsfig,graphicx,subfigure,amsfonts}
\usepackage{setspace}
\usepackage{color}
\newcommand{\be}{\begin{equation}}
\newcommand{\ee}{\end{equation}}

\newcommand{\bea}{\begin{eqnarray}}
\newcommand{\eea}{\end{eqnarray}}

\begin{document}
\preprint{APS/PRE}
\title{Liquid polymorphism, order-disorder transitions and 
anomalous behavior:\\
 a Monte Carlo study of the Bell-Lavis model for water
 }

\author{Carlos E. Fiore\footnote[1]{e-mail - fiore@fisica.ufpr.br}}
\affiliation{Departamento de F\'isica, Universidade Federal do Paran\'a, 
Caixa Postal 19044, 81531 Curitiba, PR, Brazil}

\author{Marcia M. Szortyka\footnote[2]{e-mail - marcia.szortyka@ufrgs.br}}
\affiliation{Instituto de F\'{\i}sica, Universidade Federal do Rio Grande do
Sul, Caixa Postal 15051, 91501-970, Porto Alegre, RS, Brazil}

\author{Marcia C. Barbosa\footnote[4]{e-mail - marcia.barbosa@ufrgs.br}}
\affiliation{Instituto de F\'isica, Universidade Federal do Rio Grande do
Sul, Caixa Postal 15051, 91501-970, Porto Alegre, RS, Brazil}

\author{Vera B. Henriques\footnote[3]{e-mail - vhenriques@if.usp.br}}
\affiliation{Instituto de F\'{\i}sica, Universidade de S\~ao Paulo,
Caixa Postal 66318, 05315970, S\~ao Paulo, SP, Brazil}

\date{\today}
\begin{abstract}
The Bell-Lavis model for liquid water is investigated through numerical 
simulations.
The lattice-gas model on a triangular lattice presents orientational 
states and
is known to present a highly bonded low density phase and a loosely 
bonded high
density phase. 
We show that the model liquid-liquid transition is continuous, in 
contradiction with 
mean-field results on the Husimi cactus and from the cluster 
variational method.
We define an order parameter which allows interpretation of the 
transition as 
an order-disorder transition of the bond-network. Our results 
indicate that the  
order-disorder transition is in the Ising universality class. Previous 
proposal
of an Ehrenfest second order transition is discarded. A detailed 
investigation of
anomalous properties has also been undertaken. The line of density
maxima in the HDL
phase is stabilized by fluctuations, absent in the mean-field solution.

 \end{abstract}
\maketitle

\section{Introduction}

Water is   probably the most familiar substance in nature,  due to
 its abundance and relevance for the existence of life. It exhibits 
sixty-six thermodynamic, dynamic and 
structural properties recognized to be anomalous 
\cite{URL},  that is,  unusual
 when compared with the behavior of other substances.
The most familiar anomaly  is the increase of density with temperature,
at ambient pressure, up to  $4^{\rm o}{\rm C}$. Above this
temperature water behaves as a normal liquid and density decreases as 
temperature
rises. 
Experiments for water allow to locate the line of temperatures of maximum
density (TMD), below which the
density decreases with decreasing temperature, differently from the
behavior of 
the majority of fluids, for which density increases on lowering temperature
\cite{An76}. 

In order to explain the thermodynamic anomalies, it has been proposed
that these anomalies could be related to a second critical
point at the end of a coexistence line between two liquid phases, a 
low density liquid
(LDL) and a high density liquid (HDL) \cite{Po92}. In spite of its 
experimental
inaccessibility, due to its localization beyond  the line of
homogeneous nucleation,  in the 
supercooled region, the experimental indication of the presence of 
polymorphism 
in the same region and results from numerical experiments on realistic 
water models
have maintained the idea of a second critical point alive. 

Water, however, is not 
an isolated case. There are also other examples of  tetrahedrally
bonded molecular liquids such as phosphorus \cite{Ka00,Mo03}
and amorphous  silica \cite{La00} that are other good candidates for having
two liquid phases. Moreover, other materials such as liquid metals
\cite{Cu81} and graphite \cite{To97} also exhibit thermodynamic anomalies.
Unfortunately a coherent and general interpretation of 
the low density and high density liquid phases is still missing.
Despite the lack of  consensus concerning 
the origin of water-like anomalies, 
it is widely believed that they are related to the hydrogen bond.
The tetrahedral structure of ice is a consequence of hydrogen bonding which
is responsible for ice being less
dense than the liquid phase. By increasing temperature, latent
heat is used up to disrupt hydrogen bonds which
 allows molecules to get closer. Thus 
density increases as temperature rises, up to the
temperature of maximum density (TMD).
As temperature rises further, most of the hydrogen bonds are broken and
water behaves as a normal fluid, i.e., 
the density decreases by increasing temperature.

A variety of statistical 
models have  been proposed in order to reproduce the main features of 
liquid water, and specially its anomalies. 
From a general point of view, statistical models  can be
 classified into isotropic and orientational models.
The first class of models has focalized on the 
density anomaly and its possible relation to the existence  of two
characteristic lengths with
an usual attractive interaction and a soft repulsive
interaction. The competition between these
lengths gives rise to anomalies
\cite{Ol06a,Ol06b,Wi01,He70,Ja99a,Ca03,Ku05,Ne06,Ba04,Ol05}.
Another class of isotropic  models is that in which the particles
are represented by lattice gas variables and
each particle has bonding variables represented by Potts-like states.
The
density anomaly  is introduced
\emph{ad hoc} by the addition to the free energy of a volume term proportional
to a Potts order parameter\cite{Sa93,Sa96,Fr03} .

One example of orientational model, studied both in two \cite{Pa99} and 
three \cite{Ro96} 
dimensions  also has  bonds  represented by Potts variables but
the number of bonds is limited by an energy penalty when a neighbor
site to a bond is occupied. 
This is also the mechanism for introducing 
 the density anomaly in this model.
 The second class of orientational model
emphasizes the fixed orientation
 of the hydrogen bond. One example of this type of model was introduced
by Bell and 
and collaborators 
\cite{Be70,La73,Yo79,So80} and imposes
a fixed number of possible arm states.
Further work on analogous 3-d orientational models
 revealed the possibility of a density anomaly \cite{Be94}.
More recently, Henriques and Barbosa introduced a lattice
model in two  \cite{He05a,He05b}  and
three dimensions \cite{Gi07} where water molecules have four bonding and
two inert arms (2-d) and four inert arms (3-d). The  anomalies appear due to the presence
of two competing interactions: hydrogen bonds and
 isotropic repulsive 
 van der Waals forces. In the spirit of water interactions,
the presence of nonbonding neighbors is punished by an increase of energy.

Although all models mentioned above are able to
reproduce water-like anomalies, the understanding 
of the role played by directionality is still missing. 
While in water and other tetrahedral liquids directionality 
seems to play a relevant role, it is not required to 
reproduce the thermodynamic and dynamic anomalies displayed 
in the case of isotropic potentials. It was also shown that 
the for the presence of the anomalies it is not relevant
the distinction between the
acceptor and donor arm in  the hydrogen bond 
 \cite{Ba07}.

Which would be the contribution of the directionality
in these models?  
 In order to answer this question in 
this paper we explore thoroughly the properties of  
 the Bell-Lavis model \cite{Be70}.
The model is the only 2d orientational model known to us which 
does not require an energy penalty in order to
 present a density anomaly. It is a triangular lattice gas model in 
which water molecules are represented by three symmetric bonding
arms and interact through van der Waals and hydrogen bonds.

The phase-diagram of this model has been previously
explored by means of mean-field approximations \cite{Be70,La73}, 
real space renormalization group analysis \cite{Yo79,So80} and, 
more recently, through the cluster variation method \cite{Br02} and
from Bethe calculations for the Husimi cactus \cite{Ba08}. Some Monte 
Carlo simulations
have also been presented by Patrykiejew and co-workers \cite{Pa99}.
Besides the gas-liquid transition, an open bonded network is exhibited 
by the model, at lower pressures and
temperatures (the low density liquid or open phase), which gives way to a 
dense poorly bonded network 
(the high density liquid or full phase). Consensus is lacking on the order 
of the
liquid-liquid transition. The mean-field studies predict a weak first-order 
transition, whereas
the renormalization group and the Monte Carlo simulations suggest a critical
transition. The latter also argues for a second order transition of the 
Ehrenfest type.
As to thermodynamic anomalies, these have not been investigated through 
simulations, and
Bethe calculations \cite{Ba08} have shown the density anomaly to be 
metastable. 

We also undertake a thorough investigation of the model 
thermodynamics in order
to ascertain the order of the transition, its universality class and 
the definition of 
an order parameter.  
In addition we check
for the presence of 
a stable density anomalous region of
the phase-diagram. 

This paper is organized as follows: In Sec. II the model is described
and the ground state analysis is presented, in Sec. III the simulation results
for the model thermodynamics are presented and in Sec IV final comments 
close the paper.

\section{The Bell-Lavis model and ground state analysis}
The Bell-Lavis model is a two dimensional triangular lattice gas. 
Particles are represented by occupational variables $\eta_i$, with 
$\eta_i = 0,1$ for empty or occupied sites, respectively. Each 
water particle has two orientational states, as can be seen in Fig 1a.
Orientation may be described in terms of bonding and non-bonding "arms".
The latter are represented through variables
$\tau_{i}^{ij}$ for the arm of particle $i$ which points to particle $j$. 
An arm may be non-bonding
if $\tau_{i} = 0$, or bonding, for $\tau_{i} = 1$ (see Fig. 1a and Fig. 1b).
\begin{figure}
\begin{center}
     a) \epsfig{file=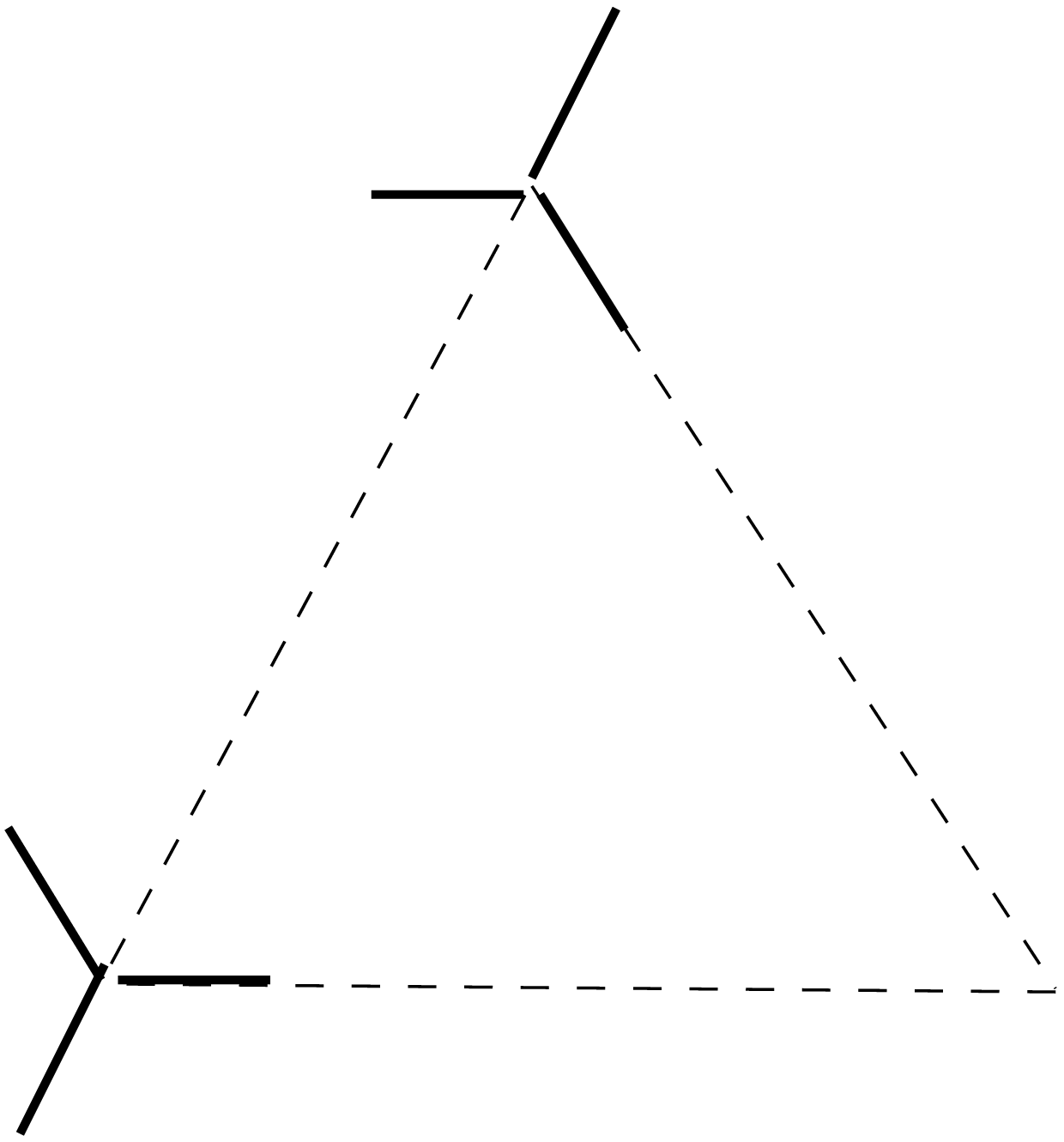,scale=0.25} 
     b) \epsfig{file=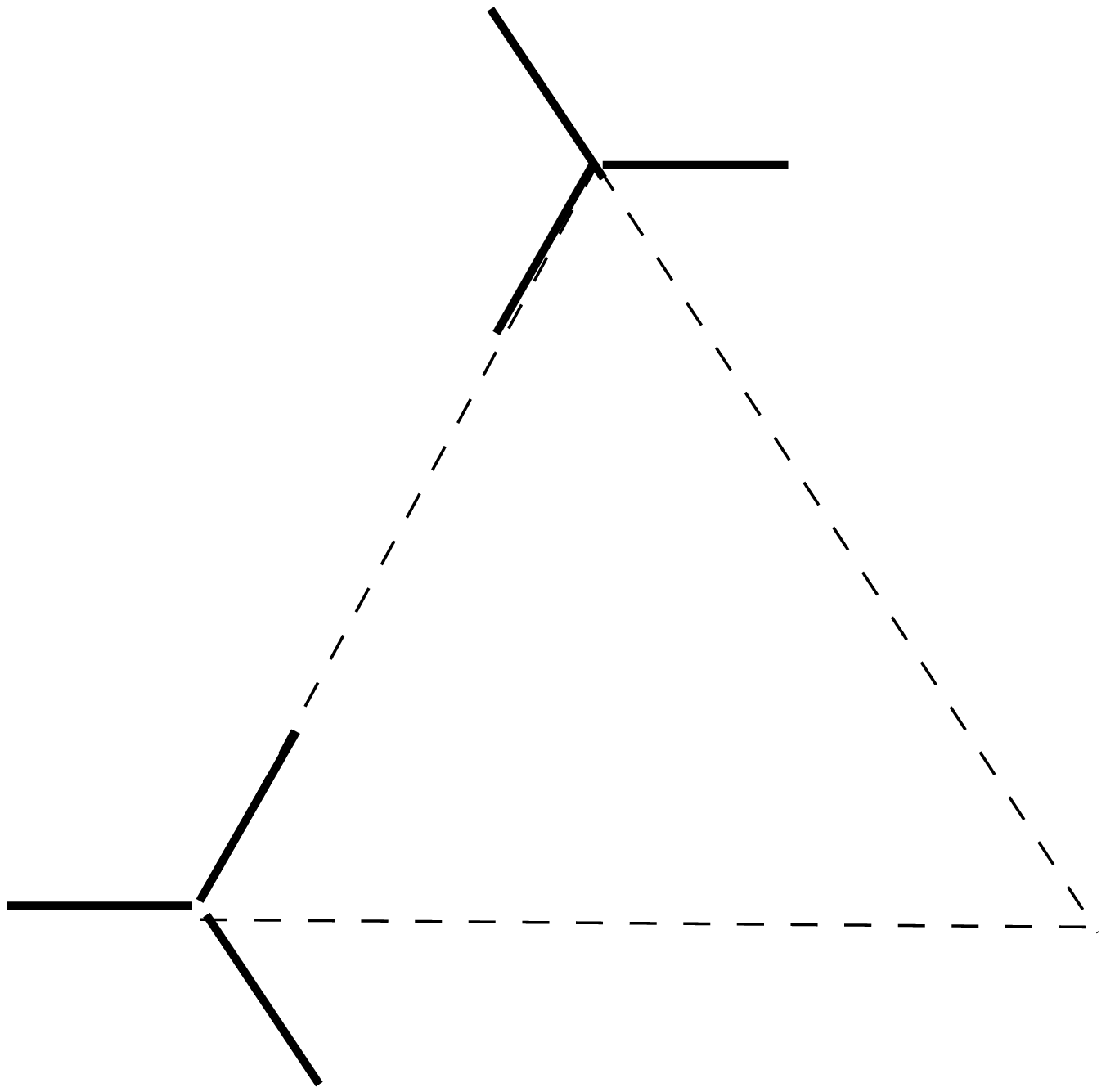,scale=0.25} 
     \caption{ \label{int} Bell-Lavis water model interactions. (a) Two 
orientations of the water particles: 
pair interaction energy is $-\epsilon_{vdw}$; (b) water molecules form a 
hydrogen bond:
pair interaction energy is $-(\epsilon_{vdw} +\epsilon_{hb})$ }
\label{fig1}
\end{center}
\end{figure}
Two next neighbor molecules are considered to interact via van der 
Waals forces and
hydrogen bonds. The model may be described by the following 
effective Hamiltonian in the grand-canonical ensemble: 
\begin{equation}
{\cal H}=-\sum_{(i,j)}
\eta_{i}\eta_{j}(\epsilon_{hb}\tau_{i}^{ij}\tau_{j}^{ij}+\epsilon_{vdw})
-\mu \sum_{i}\eta_{i},
\label{hamiltoniana}
\end{equation}
where $\epsilon_{hb}$
is the energy associated
with the formation of a hydrogen bond, in case two bonding arms point to 
each other (see Fig 1b),
$\epsilon_{vdw}$ is the van der Waals interaction energy (vdW)  and $\mu$ 
represents the chemical potential. 
The hydrogen bond interaction tends to make the particle density $\rho$ 
smaller, in order to make hydrogen bond density larger. This can be seen by 
inspection of hydrogen bonding on the lattice:
the system is fully hydrogen-bonded if the particle arms are oriented 
such as to form a honey-comb-like 
structure (see Fig 2a).
In this case, particle density is $\rho = 2/3$, and the number of
hydrogen bonds per particle is $\rho_{hb} = 3/2$.
On the other hand, the van der Waals interaction and the chemical 
potential field
tend to fill up the lattice.
However, if the lattice is fully occupied, $\rho=1$, the number of
hydrogen bonds per particle is reduced to $\rho_{hb} = 1$ (see Fig 2b).
This competition between the van der Waals and the hydrogen-bond 
interactions yields
the possibility of the appearance of 
two dense phases, with densities $\rho = 2/3$ and 
$\rho = 1$, respectively, at low temperatures.

 \begin{figure}
 
 \centering
 
 \subfigure [Example of the configuration that represents  the 
LDL phase. Note that it 
is ordered and presents, in contrast to the HDL phase,
 only one configuration.] 
 {
     \label{ldl}
     \includegraphics[clip=true,scale=0.15]{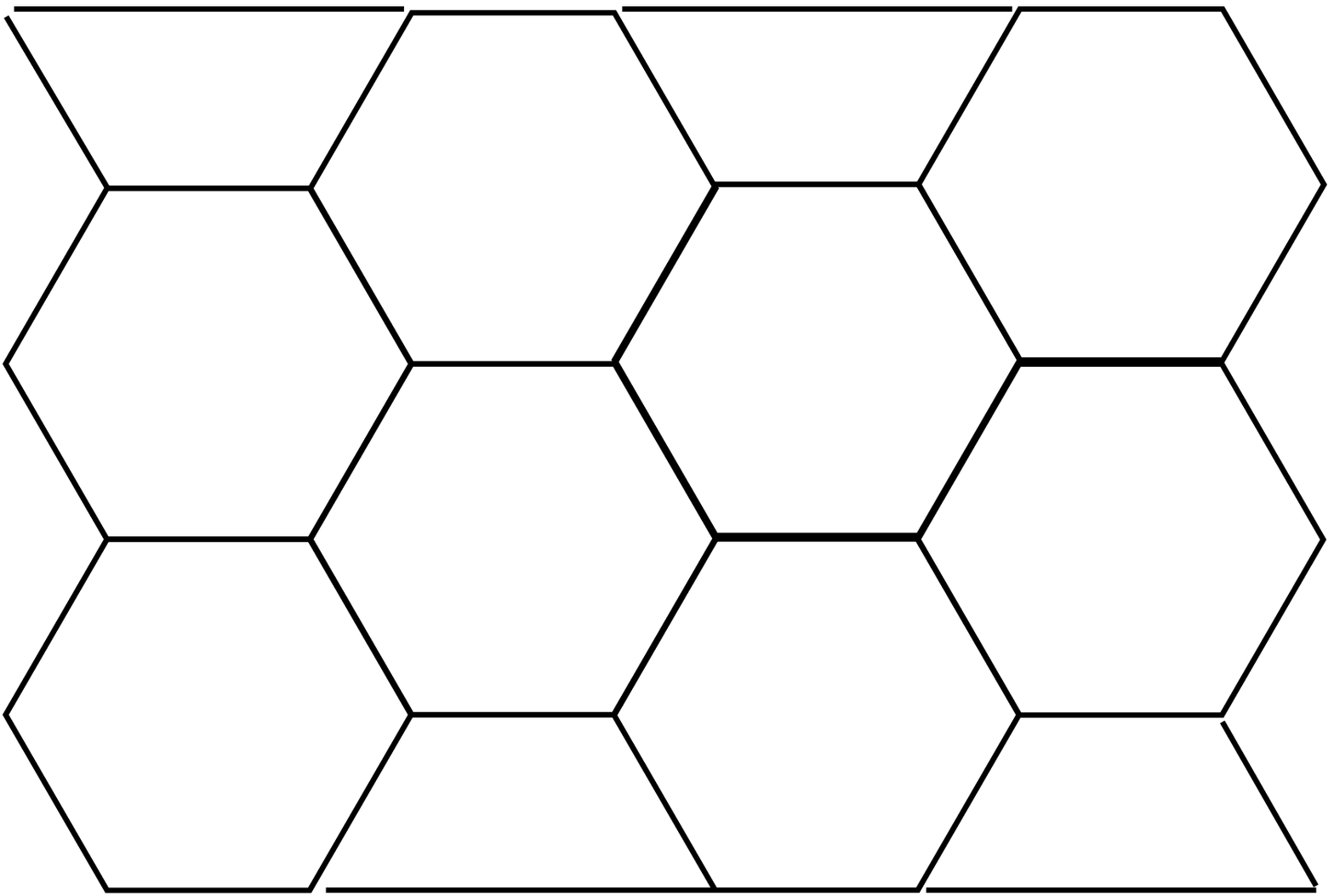}
 }
\subfigure [Example of 
a possible configuration of the HDL phase.] 
 {
     \label{hdl}
     \includegraphics[clip=true,scale=0.15]{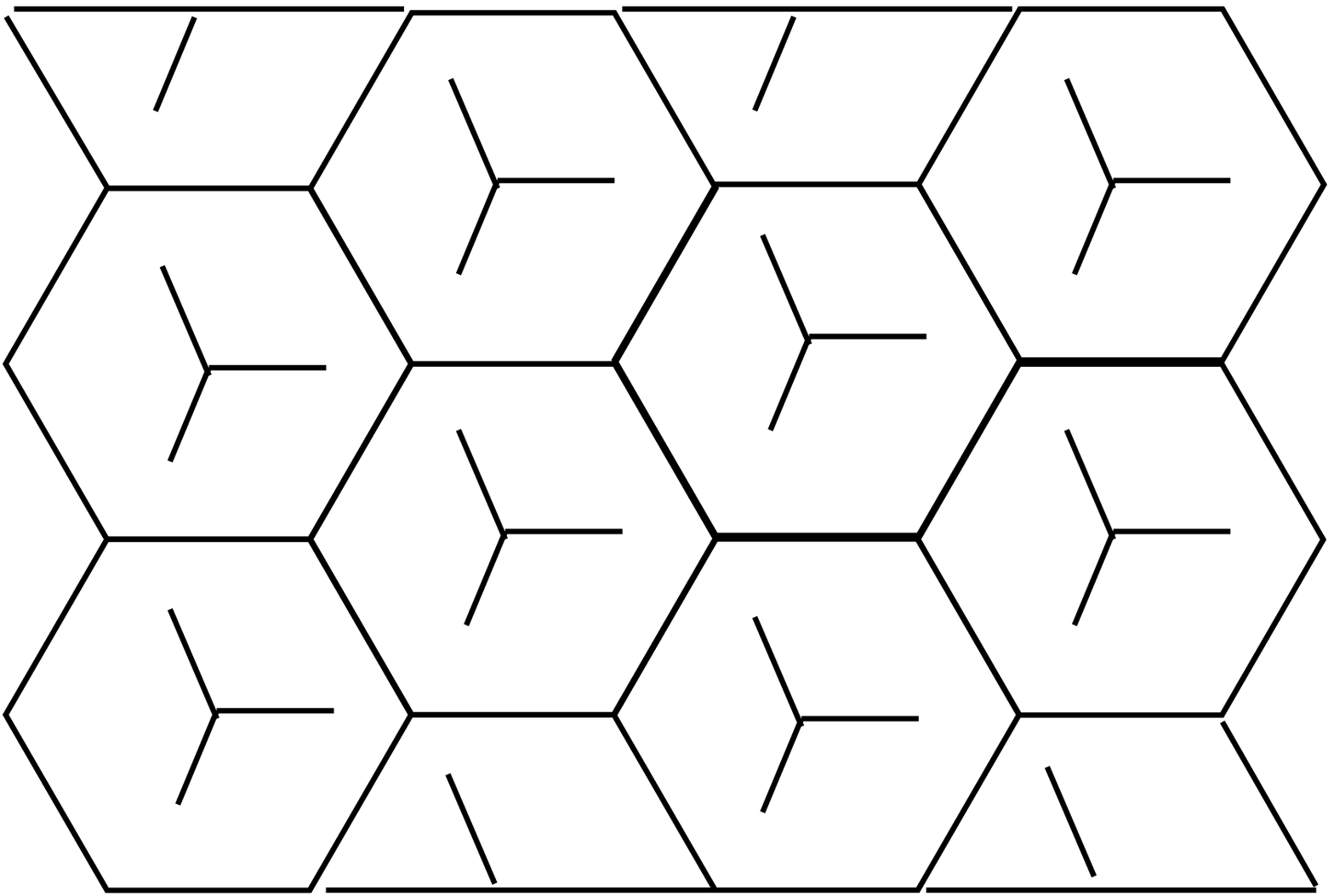}    
 }

 \caption{Liquid phases at zero temperature for the BL model.}

 \label{phases} 
 \end{figure}

The existence of two dense phases depends on the relative intensity 
of the two interactions,
$\zeta\equiv \epsilon_{vdw}/\epsilon_{hb}$. 
At zero temperature, in addition to the gas phase, with $\rho = 0$, two 
dense phases 
have been proposed for the model system, if 
$\zeta=\epsilon_{vdw}/\epsilon_{hb}<1/3$: a high density liquid 
phase, HDL, with $\rho = 1$  and a low density liquid phase, LDL, with 
$\rho = 2/3$ \cite{Pa99, Ba08}.  
The two phases are illustrated in Fig.\ref{phases} (a) and (b).

Stability of the three phases can be investigated from analysis 
of the grand potential. 
At $T=0$, the reduced grand potential free
energy density, $\phi\equiv \Phi/V\epsilon_{hb}$, is obtained 
from $\Phi=\langle \cal{H} \rangle$. 
>From inspection of the number of nearest neighbors and of 
hydrogen bonds for each phase
(see Fig. \ref{phases} ), 
one may write down the reduced grand potential for
the three possible phases as:

\begin{equation}
\phi_{gas}=0
\end{equation}

\begin{equation}
\phi_{LDL}=-1-\zeta-\frac{2}{3}\overline{\mu} \;
\end{equation}

\begin{equation}
\phi_{HDL}=-1-3\zeta-\overline{\mu};
\end{equation}
where $\overline{\mu}=\mu/\epsilon_{hb}$.

As expected, at very low and negative chemical potentials
the gas phase dominates. As the chemical
potential is increased, the low density liquid phase (LDL) becomes stable,
whereas at still larger chemical potentials the high density
liquid (HDL) corresponds to the stable phase. 

The first phase transition, between the gas and the low density liquid, 
takes place at chemical potential given by

\begin{equation}
\overline{\mu}_{\scriptscriptstyle{gas-LDL}}=-\frac{3}{2}(1+\zeta),
\end{equation}
which is obtained through equating the grand potentials, or pressures, of 
the two
phases, $\phi_{gas}=\phi_{LDL}$.
The second phase transition, between LDL and HDL, occurs
at $\phi_{LDL}=\phi_{HDL}$, which makes the chemical potential at the
transition
\begin{equation}
\overline{\mu}_{\scriptscriptstyle{LDL-HDL}} =-6\zeta.
\end{equation}
 In the next section we present our simulation studies of the model 
system properties
for finite temperatures. Our interest lies in model parameters which 
may yield a density 
anomaly, and we thus look for systems with two dense phases,
which implies taking the interaction strength of the hydrogen bond 
dominating over that
of the van der Waals parameter, i.e. $\zeta<1/3$.
We have thus focused on two cases, $\zeta=1/4$ and $\zeta=1/10$, which 
describe, respectively, weaker
and stronger hydrogen-bonding with respect to van der Waals, respectively
and which we discuss below.

\section{Thermodynamics: Monte Carlo simulations}

Model thermodynamic properties were obtained through careful
Monte Carlo simulations.  

The microscopic configurations were generated through
randomly selected exclusion, insertion or rotation of particles
in a grand-canonical ensemble, i.e., for fixed values of  temperature
and chemical potential.
Acceptance rates are those of the
usual Metropolis algorithm in the grand-canonical ensemble:
transitions between two microstates are accepted according to 
$\min\{1,\exp(-\beta \Delta {\cal H})\}$, where 
$\Delta {\cal H}$ is the effective energy difference between the two 
states. Periodic boundary conditions were adopted.
Our simulations were carried out for lattice
sizes ranging from $L=30$ to $L=90$. 

Densities are calculated as averages, as usual. Obtaining fields is a more
delicate task, in simulations.
Pressure was evaluated in two independent ways. In the first case,
the pressure was computed via numerical integration of the
Gibbs Duhem equation, $SdT-Vdp+Nd\mu=0$, at fixed temperatures, namely
$dp=\rho d\mu$.
Integration was carried out from a sufficiently low
density value, at which pressure is zero.
In the second procedure, the grand potential
free energy $\phi$  is obtained from 
the largest eigenvalue of the transfer matrix
using the Hamiltonian Eq.~(\ref{hamiltoniana}).
Since the pressure and the grand potential free energy
are related through $p=-\Phi/V$, this is an alternative which allows 
calculating the pressure directly from
the simulations and avoids performing an integration. The method
is shown in detail in the appendix A.

\subsection{Phase Diagrams: two liquids and order-disorder transition}

Phase diagrams in the
chemical potential {\it vs.} temperature plane are displayed
in Figs. \ref{fig1} and \ref{fig2} for $\zeta=1/10$ and
$\zeta=1/4$ respectively. The pressure {\it vs.} temperature phase
diagram for $\zeta=1/10$ and $\zeta=1/4$ is shown in Fig. \ref{fig3} 
and \ref{fig4}, respectively.
Reduced model parameters are used:
$\overline{T} = T/\epsilon_{hb}$, $\overline{P} = P/\epsilon_{hb}$ and 
$\overline{\mu}=\mu/\epsilon_{hb}$.
Unless otherwise stated, results presented here are for
lattice size $L=42$. 

For both 
interaction parameters analyzed, 
at low chemical potential and temperature, the system is 
constrained to the gas phase.
By increasing chemical potential 
for a fixed low temperature, a first order 
phase transition between the gas 
and the $LDL$ phase occurs. 
By increasing further the chemical potential 
at fixed low temperature  a  second order
phase transition from the $LDL$ to the 
$HDL$ phase takes place.  

At $\overline{T}=0$, the gas-LDL phase transitions take place at 
$\overline{\mu}_{\scriptscriptstyle{gas-LDL}}=-1.65$ and
$-1.875$, for $\zeta=1/10$ and $1/4$, respectively (see Eq. (5)).
As to the LDL-HDL phase transition, the corresponding points are  
$\overline{\mu}_{\scriptscriptstyle{LDL-HDL}}=-0.60$ and   
$-1.5$, respectively (see Eq. (6)).

The first-order line between the gas and the
LDL phases was investigated by means
of histograms of density, as shown in Figs. \ref{hist-0.1} and
 \ref{hist-0.25}, for smaller and larger vdW strength interactions, 
respectively.  
At phase coexistence, one has a bimodal 
distribution for the density $\rho$: the two peaks correspond to the gas 
and to the liquid densities. Figs. \ref{hist-0.1} and \ref{hist-0.25}
present the density distributions near the end of the coexistence lines,
and most probable densities are away from ground state densities 
$\rho_{gas}=0$ and $\rho_{LDL}=2/3$ and 
$\rho_{gas}=0$ and $\rho_{HDL} \approx 0.80$, respectively.
The end of first order line is characterized by a single peak in the density 
histogram, thus suggesting 
criticality. 
For $\zeta=1/10$, the gas-LDL coexistence line ends at
$\overline{T}_t = 0.435(1)$ and $\overline{\mu}_t = -1.6375(1)$. 
For $\zeta=1/4$, the
stronger van der Waals interaction extends the 
gas-liquid coexistence line to higher temperatures,
and the single peaked histogram is attained at temperature
$\overline{T}_c = 0.47$ and chemical potential $\overline{\mu}_c = -2.0095(5)$.

The  phase transition between the $LDL$ and $HDL$ phases 
presents {\it no} density 
discontinuity.  It may be identified from susceptibilities, 
as shown in Fig. \ref{fig8}
 and corresponds to a {\it second order} transition.
The terminus of the coexistence line is therefore very different in the two
cases: in the strong bond case, $\zeta = 1/10$,
 the critical LDL-HDL line meets the gas-LDL coexistence 
line at a tricritical point (TCP),
whereas for the weak bond case, $\zeta=1/4$, the critical HDL-LDL line 
meets  the coexistence  line at a critical endpoint (CE). 
In the latter case, the gas-liquid coexistence line ends at a critical point.
\begin{figure}[h!]
\includegraphics[clip=true,scale=0.4]{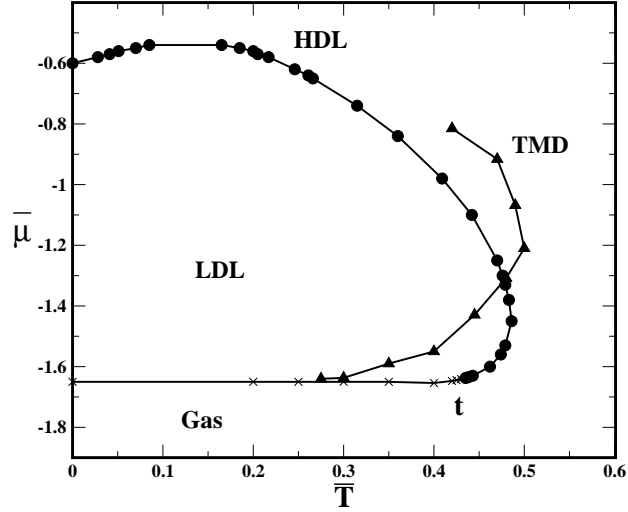}
\caption{Phase diagram for the Bell-Lavis model in the space
of reduced chemical potential $\overline{\mu}$ vs. reduced temperature 
$\overline{T}$ for $\zeta=1/10$.
Stars, circles and triangles denote the phase transition between 
gas-$LDL$, $LDL-HDL$ phases and TMD line, respectively.}
\label{fig1}
\end{figure}
\begin{figure}[h!]
\includegraphics[clip=true,scale=0.4]{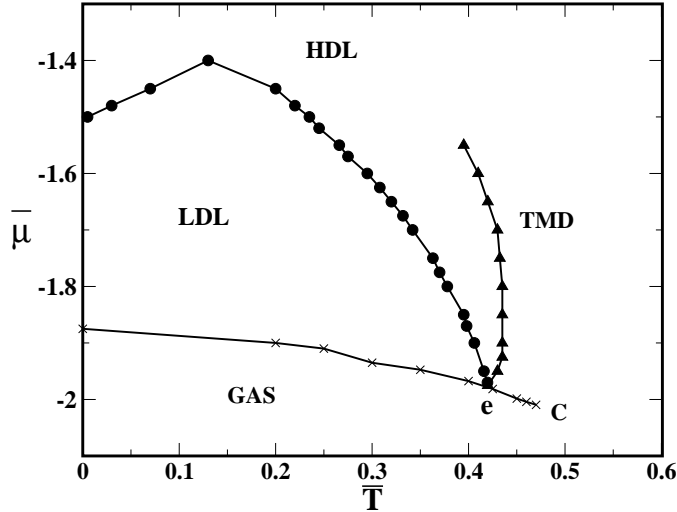}
\caption{Phase diagram for the Bell-Lavis model in the space
of reduced chemical potential $\overline{\mu}$ 
versus reduced temperature $\overline{T}$ for $\zeta=1/4$.
Stars, circles and triangles denote the phase transition between 
gas-$LDL$, $LDL-HDL$ phases and TMD line, respectively.}
\label{fig2}
\end{figure}
\begin{figure}
\includegraphics[clip=true,scale=0.4]{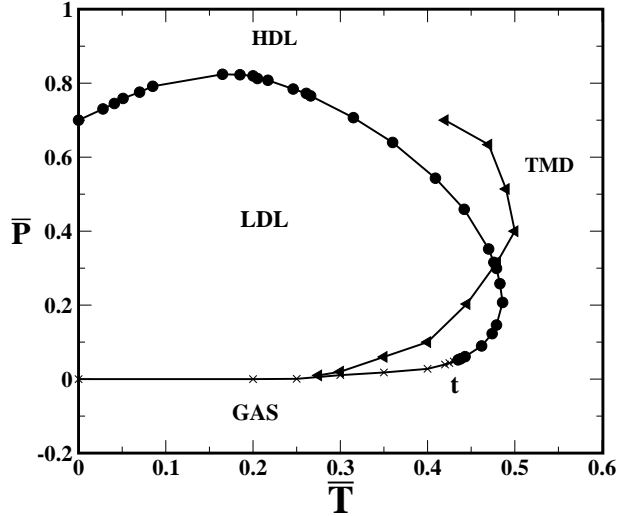}
\caption{Phase diagram for the Bell-Lavis model in the space
 of reduced pressure $\overline{P}$ vs. reduced temperature 
$\overline{T}$ for $\zeta=1/10$. 
 Stars denotes the first-order phase transition 
between the  gas-$LDL$, circles the continuous second-order 
phase transition between the $LDL-HDL$ phases and triangles denotes the 
TMD. The first-order phase transition 
between  gas-$LDL$ meets the continuous transition at the tricritical
point $t$.}
\label{fig3}
\end{figure}
\begin{figure}
\includegraphics[clip=true,scale=0.4]{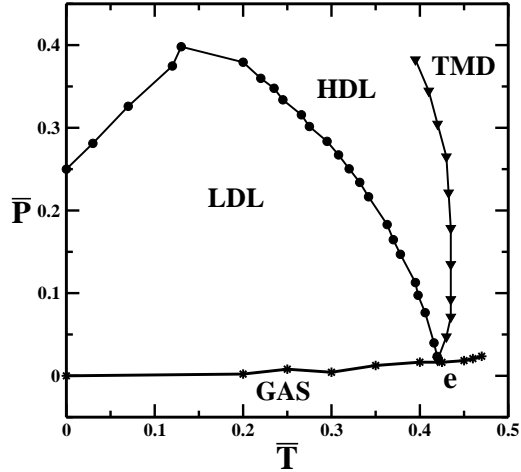}
\caption{Phase diagram for the Bell-Lavis model in the space
 of reduced pressure $\overline{P}$ vs. reduced temperature 
$\overline{T}$ for $\zeta=1/4$. 
 Stars denotes the first-order phase transition 
between  gas-$LDL$, circles the continuous second-order 
phase transition between the $LDL-HDL$ phases and triangles denotes the 
TMD. The continuous phase transition 
between  $LDL-HDL$ phases  ends at the 
first-order phase boundary between the  gas-$LDL$
at a critical endpoint $e$.}
\label{fig4}
\end{figure}
\begin{figure}
\includegraphics[clip=true,scale=0.4]{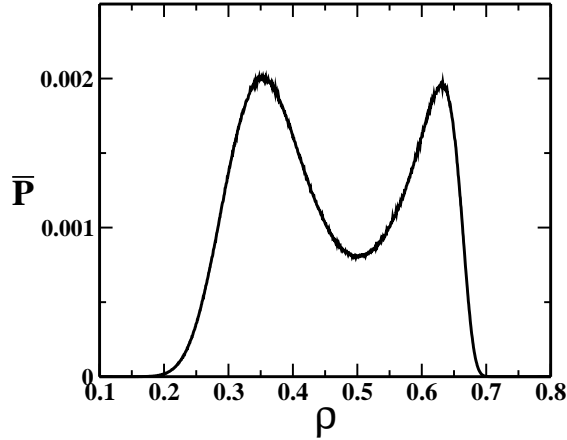}
\caption{Histogram of the density of molecules $\rho$ for the first 
order line between the gas and $LDL$ phases for $\zeta = 1/10$, 
 $\overline{\mu}=-1.6472(3)$ and $\overline{T}=0.42$.}
\label{hist-0.1}
\end{figure}
\begin{figure}
\includegraphics[clip=true,scale=0.4]{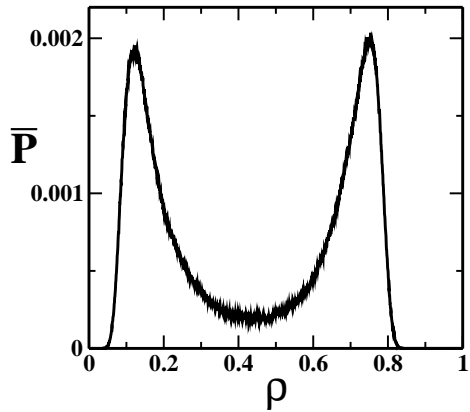}
\caption{Histogram of the density of molecules 
$\rho$ for the coexistence
 phase between the gas and $HDL$ phases 
for  $\zeta=1/4$, $\overline{\mu}=-1.99850(3)$ 
and $\overline{T}=0.45$.}
\label{hist-0.25}
\end{figure}

The differences between the phase diagrams can be rationalized as 
follows: stronger 
bonds relative to van der Waals interactions, $\zeta=1/10$, lead to
a larger LDL phase, whereas 
stronger van der Waals interactions with respect to bond 
interactions, $\zeta=1/4$, stabilize 
gas-liquid coexistence at higher temperatures. Extension 
of the liquid-gas coexistence line
together with contraction of the LDL phase transform the 
tricritical point into a critical end point.

It may also be noted that the phase diagram displays 
reentrant behavior of the LDL-HDL line:
the HDL phase is the 
lower entropy phase at low temperatures, whereas at
higher temperatures, the LDL phase becomes of lower entropy 
with respect to the HDL phase. 

Our phase diagrams must be compared to some results present in the 
literature. 
The exact solution on a Husimi cactus \cite{Ba08} for the same model 
parameters, $\zeta=1/10$ and $\zeta=1/4$
has yielded weak first-order LDL-HDL transitions. A previous study 
by Bruscolini et al. \cite{Br02} 
with the cluster variational method of the $\zeta=1/4$ case 
led to the same conclusion.
On the other hand, Patrykiejew and co-workers \cite{Pa99} obtain 
through Monte Carlo simulations a continuous 
liquid-liquid transition line for the $\zeta=1/4$ case, in 
accordance with our results. 
Thus the first-order liquid-liquid transition seems to be an 
artifact of the Bethe-like
solutions.

However, on a global look, the Husimi cactus solution \cite{Ba08} 
produces phase diagrams qualitatively similar
to our own: for the stronger bond $\zeta=1/10$ case, the 
critical point, present for the weaker
bond case, $\zeta=1/4$, disappears, and the gas-liquid line 
joins smoothly the liquid-liquid line.

\subsection{Two liquids and order-disorder transition}

Now a question may be posed: the absence of a density  gap 
indicates that the model {\it does not} display  
liquid-liquid coexistence, so  what distinguishes the two phases?

\begin{figure}
\begin{centering}
\includegraphics[clip=true,scale=0.6]{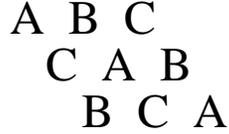} 
\par\end{centering}
\caption{Three sublattices on the triangular lattice, named A B C.}
\label{sublattice} 
\end{figure}

A previous study on the mapping of the BL model on an 
anisotropic spin-one antiferromagnetic model 
\cite{Ba08} suggested sublattice ordering, corresponding
 to non-frustrated antiferromagnetic
ordering on the triangular lattice. We have thus examined 
the model sublattice
properties. In order to proceed, we have divided the triangular lattice 
into  three sublattices 
named $A$, $B$ and $C$, as illustrated in Fig.(\ref{sublattice}). 
We have measured sublattice average density and molecular orientational 
state.

\begin{figure}
\includegraphics[scale=0.35]{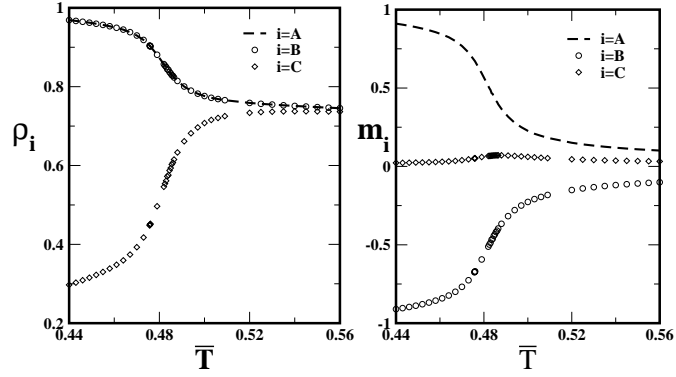}
\caption{Graph $(a)$ presents the plot of sublattices densities $\rho_{i}$,
 with $i=A, B, C$ {\it vs} reduced temperature
$\overline{T}$. In graph $(b)$, we have average particle orientation
$m_{i}$ on each sublattice {\it vs} $\overline{T}$. 
In both cases, $\mu=-1.40$ and 
$\zeta=1/10$.}
\label{fig6}
\end{figure}

In Fig. \ref{fig6}(a), we plot the  density  per site $\rho_i$ 
on each sublattice, for low strength van der Waals, $\zeta=1/10$.
It can be seen that as temperature is lowered 
two sublattices (A and B) are filled with particles while
the third sublattice (C) becomes empty. This occurs rather abruptly in
the same range of temperatures of the specific heat peak ($\overline{T} 
\approx 0.5$).
This suggests using an order parameter $\psi$ given by 
\begin{equation}                                                               
\psi=\rho_i - \rho_j,
\label{e100}                       
\end{equation}
with $i,j=A,B$ or $C$.
At $\overline{T}=0$, $|\psi|=1$ or $0$, depending on the pair of sublattices 
chosen, whereas 
at high temperatures, $\psi=0$, for any two pair of sublattices.

We have also compared molecular orientation on sublattices.
We call the two orientational states presented in Fig. 1(a) as $m=-1$
and $m=+1$ states, respectively. 
Fig. \ref{fig6}(b) presents  the
average value of the variable $m$ as a function of temperature, for each
sublattice. At low temperatures, the high-density sublattice A presents
molecules in one of the orientational states, $m\approx +1$, 
whereas the second high-density sublattice, B, 
presents particles mostly in the opposite orientational state,
with $m\approx -1$. In the low density sublattice, C, molecules present
no preferential orientation, and we have $m\approx 0$.
As the specific heat 
peak position ($\overline{T}\approx 0.5$) is approached, 
molecule average orientation becomes random. Therefore,
the LDL-HDL transition may then be characterized as an order-disorder
transition. Positional as well as orientational order disappear at the 
transition.
 
\subsection{Critical line}

\begin{figure}
\includegraphics[clip=true,scale=0.35]{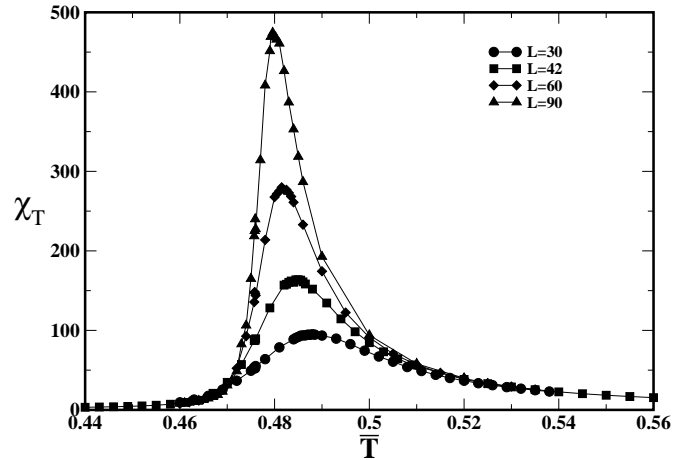}
\caption{Isothermal susceptibility $\chi_{\scriptscriptstyle{T}}$ 
{\it vs.} reduced temperature $\overline{T}$ for 
$\zeta=1/10$ and
$\overline{\mu}=-1.40$ and different lattice sizes $L$.}
\label{fig8}
\end{figure}

In order to give a more precise definition of the continuous 
order-disorder transition line,
the second moment of the order parameter 
$\psi$ has been investigated.
We have computed
the  isothermal susceptibility given by 
\begin{equation}
\chi_{\scriptscriptstyle{T}}=\frac{V}{T}(\langle \psi^{2}\rangle -\langle \psi \rangle^{2}).
\end{equation}
This susceptibility is shown in Fig. \ref{fig8} as a function 
of temperature, for $\zeta=1/10$, at $\overline{\mu}=-1.40$.
The peak in the 
susceptibility grows with $L$, suggesting that
the system undergoes a 
phase transition at
$\overline{\mu}=-1.40$ and $\overline{T} \approx 0.48$.
Analogous measurements for different chemical potentials
were undertaken in order to build the LDL-HDL transition
line in the phase diagram of Figs. \ref{fig1} and \ref{fig2}.
The corresponding line is a line of susceptibility maxima. 

In order to check the location  of the
critical line, the fourth-order cumulant for the order parameter
\begin{equation}
U_{4}=1-\frac{\langle \psi^{4}\rangle}{3\langle \psi^{2}\rangle^{2}}\;,
\end{equation}
was computed 
for different lattice sizes. The results are shown in Fig. \ref{fig10}
for $\zeta=1/10$, $\overline{\mu}=-1.40$ and lattice sizes 
$L=30, 42, 60, 90$.
The crossing of the lines representing different lattice sizes 
at a single point
confirm the presence of criticality. Computed cumulants for other
values of the chemical potential display analogous behavior,
lending confidence to the interpretation of criticality. 

But to what
universality class does the critical order-disorder line belong?
In order to answer this question we have
analyzed the value of the cumulant at the crossing point as well as
the order parameter scaling with size.

The cumulant $U_{4}$ displays different regimes at low and high temperatures:
for low temperatures, it approaches the 
value $2/3$, while for high temperatures 
it approaches $0$. At the phase transition temperature, 
 $\overline{T}_{c}=0.4760(2)$,  
it displays a non trivial value $0.610(5)$ for all lattice sizes. 
The non-trivial  value of the  cumulant $U_{4} \approx 0.610$ 
at the criticality is characteristic
 of systems belonging to the Ising universality class.

\begin{figure}
\includegraphics[scale=0.3]{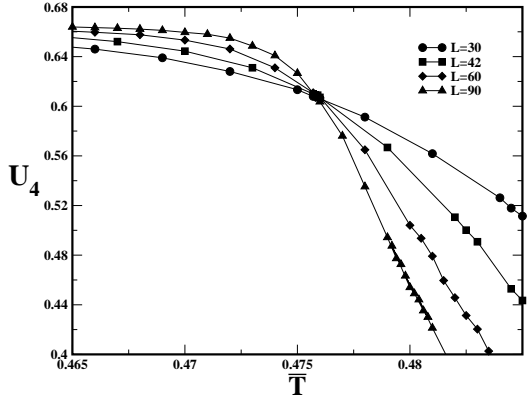}
\caption{Fourth-order cumulant versus reduced temperature 
$\overline{T}$ for
 $\overline{\mu}=-1.40$, $\zeta = 1/10$ and different lattice sizes $L$. }
\label{fig10}
\end{figure}

\begin{figure}
\includegraphics[scale=0.3]{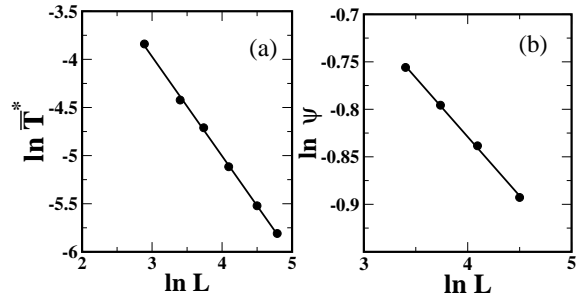}
\caption{In the graphs $(a)$ and $(b)$ we have a log-log plot of
the  $\overline{T}^{*} \equiv 
\overline{T}_{L}-\overline{T}_{\infty}$ and $\psi$ vs. $L$
for $\zeta=1/10$ and $\overline\mu=-1.40$.
The continuous lines have slope $1.03(2)$ and $0.124(3)$, respectively. }
\label{fig9}
\end{figure}

Next we examine the scaling of the order parameter.
According to the finite size scaling theory 
\cite{Bi92},  at the critical point the order parameter
decreases algebraically with the system size
through the relation $\psi \sim L^{-\beta/\nu}$, where 
$\beta/\nu$ is the associated critical exponent. The critical 
exponent $\nu$ describes 
the spatial length correlation $\xi$ which diverges at the critical 
point according to the law $\xi \sim {\bar t}^{-\nu}$, 
where ${\bar t}= \overline{T}- \overline{T}_{c}$. For
finite systems, it leads to the expression 
 $\overline{T}_{L}-\overline{T}''_{c}\sim L^{-1/\nu}$ \cite{Bi92}, where 
 $\overline{T}_{L}$ is the pseudo critical temperature, obtained by a
 maximum in ``some susceptibility''.
 Therefore, log-log plots of $\psi$ and 
$\overline{T}_{L}-\overline{T}_{\infty}$ {\it vs.} $L$ yield the exponents 
$\beta/\nu$ and $\nu$,
respectively. Fig. \ref{fig9} (a) and (b) illustrate such plots for
 $\zeta=1/10$
and $\overline{\mu}=-1.40$. From the plots we obtain
 $\beta/\nu = 0.124(3)$ and $\nu = 1.03(2)$. These values are in excellent 
agreement with exact values for
the Ising model $\beta=1/8$ and $\nu=1$, thus classifying the 
order-disorder transition
of the BL model in the Ising universality class.

\begin{figure}
\setlength{\unitlength}{1.0cm}                                               
\includegraphics[scale=0.35]{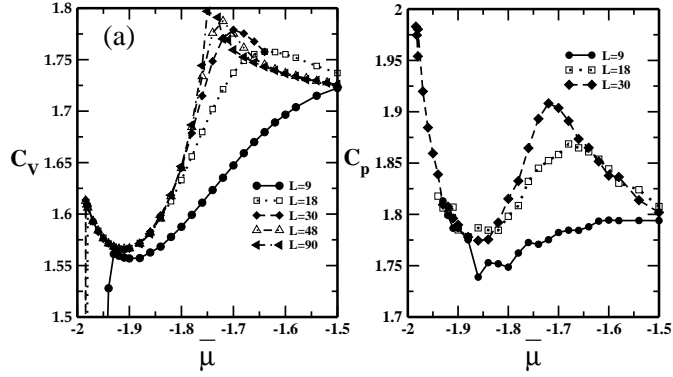}
\caption{Specific heat at constant volume $c_V$ and at
constant pressure  $c_P$ versus
$\overline \mu$ for $\overline T=0.35$ for the BL model
and $\zeta=1/4$. }
\label{fig11}
\end{figure}

This conclusion is in contrast to the suggestion of Patrykiejew 
\cite{Pa99} and co-workers
for the continuous liquid-liquid line. They propose that it would 
be an example of a  
second-order phase transition in the Ehrenfest classification,
as demonstrated by discontinuity of the specific heat at constant 
volume $C_V$ at the
transition. 
In Fig. \ref{fig11}  we show the dependence of $c_V=C_v/V$ and 
$c_P=C_P/V$ versus the reduced chemical potential 
$ \overline\mu$  for $\overline T=0.35$. The constant volume and
constant pressure specific heats were calculated from simulation data
at constant chemical potential through expressions \cite{Al87}
\begin{equation}
C_V=\frac{1}{k_{B}T^{2}}(\langle \delta {\cal H}^{2}\rangle_{\mu V T}-
\frac{\langle \delta {\cal H} \delta N\rangle^{2}_{\mu V T}}{\langle 
\delta N^{2}\rangle_{\mu V T}}),
\end{equation}
and 
\begin{equation}
C_P=C_V+TV\alpha_{P}^{2}/{k_{T}},
\end{equation} 
where $k_{T}=K_{T}/V$, 
 $K_{T}=\frac{V}{N^{2} k_{B} T}\langle \delta N^{2}\rangle_{\mu V T}$ and 
\begin{equation}
\alpha_{P}=\frac{P K_{T}}{T}-
\frac{\langle \delta {\cal H} \delta N\rangle_{\mu V T}}{N k_{B}T^{2}}+
\frac{\langle {\cal H} \rangle_{\mu V T} 
\langle \delta N^{2}\rangle_{\mu V T}}{N^{2}k_{B}T^{2}},
\end{equation}
where $N=\sum_{i=1}^{V}\eta_{i}$ and $\delta X=X-\langle X \rangle$ with 
$X={\cal H}$ and $N$.
Our results show that the constant volume specific heat $c_V$ displays 
a discontinuity
at $\overline \mu=-1.98$, close to the gas-liquid transition line 
(see Fig. \ref{fig4}), 
and a small peak close to $\overline \mu=-1.74$, that increases 
by increasing $L$,  which is  in consistency
with the transition point in the corresponding phase diagram 
(Fig. {\ref{fig4}})
obtained by means of the 
isothermal susceptibility analysis. In Ref.\cite{Pa99} the authors might
 have been misled by the
absence of a phase diagram in the chemical potential vs. temperature 
plane. The specific heat
presented in Ref.\cite{Pa99} corresponds
to our temperature $\overline T=0.175$ \cite{footnote}.
At this temperature, the gas-liquid transition is near 
$\overline \mu=-7.4$, in their units, whereas
the liquid-liquid transition is near $\overline \mu=-5.7$, in the 
same units. The discontinuity presented
in the paper is near $\overline \mu=-7.3$, and thus must correspond 
to the gas-liquid transition.
Ranges below $\overline \mu=-6$ are absent from their figure, so the 
liquid-liquid transition peak
is not shown.

\subsection{Anomalous Properties}
 
In this section we present data for the model particle and H-bond densities, 
and for model entropy. 

The presence of a low density liquid suggests that
a line of maxima of densities exists. Such maxima were looked for both at
constant chemical potential and constant pressure and displayed in 
the model phase diagrams as TMD lines (see Figs. \ref{fig1} and \ref{fig2}).
Note that the TMD crosses the LDL-HDL critical line in the case of strong
bonds ($\zeta=1/10$): the anomaly is inside the LDL phase at low pressures 
and migrates to the HDL phase at higher pressures. In the case of weaker bonds 
($\zeta=1/4$), the 
anomaly is present only in the HDL phase. 

However, as discussed in the previous subsection, correlations between
system density and hydrogen-bond density per particle seem to be of some
importance. We therefore compare the
behavior of the two densities with temperature, for both
strong and weak bonds. 
Figs. \ref{fig10} (a) and (b) show data for the density $\rho$ {\it vs.}
 $\overline T$, for different fixed pressures $P$. For low pressures, 
the density presents a maximum. In contrast, for
higher pressures $P$, the density is a decreasing function of temperature.
Particle density behavior is closely accompanied by hydrogen-bond
density behavior (Fig. \ref{fig10} (c) and (d)). For the lower pressures, 
for which densities present a 
maximum, hydrogen bond densities decrease with temperature. 
For the lowest pressure, at which a density maximum is clearly seen, 
an inflection point of the H-bond density occurs is present at
the same temperature.
On the other hand, for the higher pressures, for which density is a decreasing
function of temperature, hydrogen bond densities increase mildly,
at low temperatures. The low pressure behavior coincides with the
qualitative picture which has been ascribed to water for a long time:
density increases while the hydrogen-bond network distorts, implying
a decreasing H-bond density. On the other hand, the increasing H-bond
density at higher pressures, low temperatures, suggests that the appearance
of empty sites allows for more bonding.

\begin{figure}
\setlength{\unitlength}{1.0cm}
\includegraphics[scale=0.47]{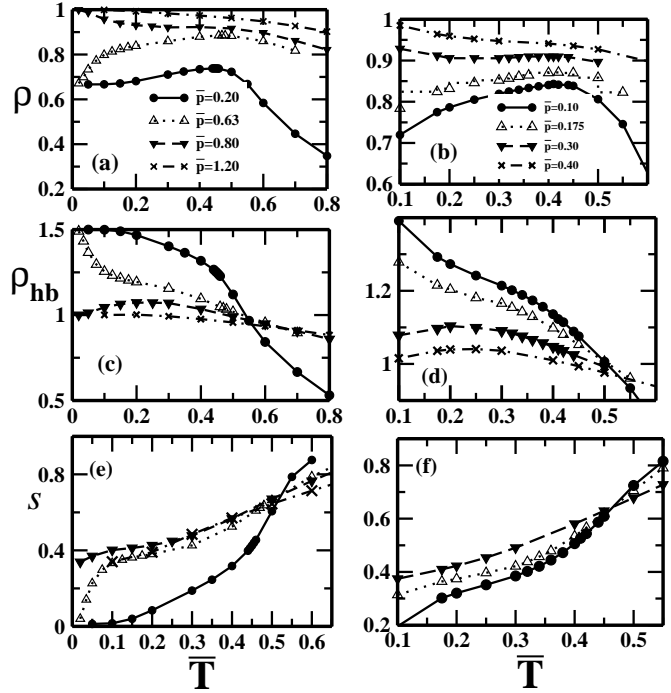}
\caption{Dependence of the quantities  $\rho$, $\rho_{hb}$ and 
$s$ versus $\overline T$ for several
values of pressure $p$ for the BL model. The graphs $(a),(c), (e)$
and $(b),(d), (f)$
restrict  to the regimes $\zeta=1/10$ and $\zeta=1/4$, respectively. }
\label{fig10}
\end{figure}

Finally, entropy per site is shown in Fig. \ref{fig10} (e) and (f).
For the strong bond case, $\zeta = 1/10$, the degeneracy of the
HDL ground state is clearly seen (for $\overline P=0.8$ and $\overline P=1.2$).
Moreover, the thermodynamic identity $\delta S/\delta P=-\delta V/\delta T$
allows interpretation of entropy behavior as complementary to density
behavior (Fig. \ref{fig10} (a) and (b)). Note the inversion of the relative
position of entropies at different pressures, at fixed temperature,
before and after the curves cross: at low temperatures, the
low pressure curves, which present increasing density as a function
of temperature, display lower entropy than the higher pressure entropies,
associated with monotonic decreasing density, as function of temperature.
At higher temperatures, the situation is opposite: the lower pressure entropies
are higher than the higher pressure entropies. Thus the anomalous density
behavior is accompanied by an "anomalous" entropy behavior, in which
entropy increases with pressure. Differently from normal liquids, this
can be rationalized in terms of disordering of bonds: entropy increases
because bond disordering dominates over positional ordering, as density
is increased with pressure.


\section{Final Comments}

We have investigated the Bell-Lavis model for liquid water through 
numerical simulations in order to shade some light in the 
role played by the orientational degrees of freedom of this
model in the liquid-liquid transition.
Our study has allowed the clarification of the nature of the 
liquid-liquid transition.
Previous careful mean-field studies, such as calculations on the 
Bethe lattice \cite{Ba08} or through the
cluster variation method \cite{Br02} yielded liquid-liquid 
coexistence with a small density gap. 
The Monte Carlo simulations demonstrate that the transition is 
continuous. Moreover,
our finite scaling analyses indicate that the transition is in the 
Ising universality class.


In the absence of a density gap, characterization of the transition requires
establishing an order parameter. Inspired on the mapping on the antiferromagnetic
spin model, we propose an order parameter based on the difference in sublattice densities,
associated to the highly bonded configurations. This order parameter presents a divergent
susceptibility at the critical temperature. On the other hand, we have also shown that
positional order on sublattices is accompanied
by orientational order. Thus the ordered LDL phase presents both positional and orientational order
which disappear in the HDL phase.

In the analysis of the thermodynamic variables we were able to accompany number and H-bond densities
as well as entropy per particle. The latter is calculated directly from simulations through a transfer matrix
representation of the model Hamiltonian. Comparison of the behavior of the three densities with temperature
shows that the density anomaly is accompanied by inflection
points both in hydrogen bond density as well as in entropy per particle. Such behavior was suggested in the
mean-field approach \cite{Ba08}, but is made much more clear in the simulation data.

In relation to the density anomaly, it must be pointed out that the line of density maxima (TMD) which was 
located in the metastable HDL phase for the Bethe lattice, is turned stable through fluctuations present in 
the Monte Carlo procedure.

As a final remark, we should say that a real liquid-liquid transition would not be continuous transition, since liquid 
polymorphism is understood do imply discontinuity in density across the transition. Could
the transition become first-order in three dimensions? This is a point to be cleared. However,
the Bell-Lavis model has no trivial extension to three dimensions. Nonetheless, the feature that makes it
an interesting model is the fact that it is an orientational model with attractive van der Waals
interactions. This is not the case for every other orientational model in the literature that we know of. 
Thus it remains to be cleared whether orientational models with attractive isotropic interactions
are able to yield liquid-liquid coexistence.

\subsection*{Acknowledgments}

C.E.F acknowledges the partial financial support by 
FAPESP, under grant 06/51286-8 and CNPq.

\appendix

\section{Pressure Calculation}

In this appendix the pressure is obtained from the grand  potential free
energy \cite{Sa95}.
In order to describe the method briefly, let us 
consider a triangular lattice with $V$ sites divided in $N$ successive
layers $S_i \equiv (\eta_{1,i},\eta_{2,i},...,\eta_{N,i})$ with $L$
sites, $V = L \times N$. The Hamiltonian may be decomposed in the following way
\begin{equation}  
{\cal H}= \sum_{k=1}^N {\cal H}(S_k,S_{k+1}),\label{e14}                       \end{equation}
where due to the periodic boundary conditions $S_{N+1} = S_1$.
                                                                 
The probability $P(S_{1},S_{2},...,S_{N})$ of a given configuration
of the system is given by
\begin{equation}                                                               P(S_{1},S_{2},...,S_{N})=\frac{1}{\Xi}
 T(S_{1},S_{2}) T(S_{2},S_{3})...T(S_{N},S_{1}),                                \label{e15}                   
\end{equation}
where $T(S_{k},S_{k+1}) \equiv \exp ( - \beta {\cal H}(S_k,S_{k+1}))$ 
is an element of the transfer matrix $T$ and
\begin{equation}                    
\Xi=\rm Tr(\it {T^{N}}),\label{e16}                                              \end{equation}
where $\Xi$ is the Grand-Canonical partition function. By using the spectral
expansion of the matrix $T$ it is possible to show \cite{Sa95} that
\begin{equation}                            
\lambda_0=\frac{ <T(S_{1},S_{1})>}{<\delta_{S_{1},S_{2}}>},
 \label{e17}                                               
\end{equation}
where  $\lambda_0$ denotes the largest eigenvalue, which  is evaluated
from averages $<T(S_{1},S_{1})>$ and $<\delta_{S_{1},S_{2}}>$.
The quantity $T(S_{1},S_{1})$ is obtained from $T(S_{k},S_{k+1})$
by taking $S_k=S_{k+1}$, where 
\begin{eqnarray}
T(S_{k},S_{k+1})=\exp \{\sum_{i=1}^{L}[
 \eta_{i,k}(\eta_{i,k+1}+\eta_{i+1,k}
+\eta_{i+1,k+1})\\(\epsilon_{vdw}+\epsilon_{hb}
 \tau_{i,k}(\tau_{i,k+1}
+\tau_{i+1,k}+\tau_{i+1,k+1})+ \mu \eta_{i,k})] \},
\nonumber
\end{eqnarray}
and $\delta_{S_{1},S_{2}} = 1$ when layers $S_{1}$ and $S_{2}$ are
equal and zero otherwise. 
Finally, the free energy is evaluated from the largest eigenvalue 
through the relation
\begin{equation}                                                               
\frac{\Phi}{V}=-\frac{1}{\beta V} \ln \lambda_{0} = - P   
\label{e28}                       
\end{equation}
where $V$ is the volume (number of sites in the lattice) and 
$\Phi$ is the grand potential free energy.
The entropy per site is evaluated from the grand potential through the
 formula 
\begin{equation}
s=\frac{u-\phi}{T},
\end{equation}
where $u=U/V$ and $\phi=\Phi/V$.

\bibliographystyle{aip}
\bibliography{Biblioteca}

\begin{thebibliography}{10}
\bibitem{footnote}
all their reduced values have a factor $4$, since they have considered reduced
  unities in terms of the van der Waals energy $\epsilon_{vdW}$ .
\end{thebibliography}

\end{document}